\documentclass[a4paper,12pt]{article}
\usepackage{amssymb}
\usepackage{graphicx}
\usepackage{hyperref}
\usepackage{harvard}
\usepackage{amsmath}
\usepackage[english]{babel}
\usepackage{amsfonts}
\usepackage{rotating}
\usepackage{lscape}
\usepackage[official]{eurosym}
\usepackage{longtable}
\usepackage[a4paper,left=2.5cm,right=2.5cm,top=3cm,bottom=3cm]{geometry}
\usepackage{xspace}
\usepackage{array}
\usepackage{latexsym}
\usepackage{setspace}
\usepackage{xspace}
\usepackage{authblk}
\usepackage{rotating}
\usepackage{tabularx}
\usepackage{placeins}
\usepackage{multirow}
\usepackage{pdfpages}
\usepackage{xcolor}
\usepackage{booktabs}
\usepackage{authblk}

\begin{document}

%
%
%
%

\title{{\huge Gender Differences in Wage Expectations}}

\author{
Ana Fernandes \thanks{Berner Fachhochschule, Br{\"u}ckenstrasse 73. 3006 Bern, Switzerland; ana.fernandes@bfh.ch}   \hskip 1cm
Martin Huber \thanks{University of Fribourg, Bd.\ de P\'{e}rolles 90, 1700 Fribourg, Switzerland; martin.huber@unifr.ch. } \hskip 1cm
Giannina Vaccaro \thanks{Vaccaro is very grateful for financial support from the Swiss National Science Foundation (SNSF); gianninavaccaro@outlook.com.}
\thanks{We have benefitted from comments by conference participants at the 9\textsuperscript{th} ifo Dresden Workshop on Labor Economics and Social Policy, the 2019 Annual Congress of the Peruvian Economic Association, and the 2019 Annual Congress of the Swiss Society for Economics and Statistics.}
}


%


%

\date{}
\maketitle

\begin{center}


\date{\today}

\end{center}

\begin{center}\textbf{Abstract} \end{center}
{
	
\renewcommand{\baselinestretch}{1.0}
	\noindent
	Using a survey on wage expectations among students at two Swiss institutions of higher education, we examine the wage expectations of our respondents along two main lines. First, we investigate the rationality of wage expectations by comparing average expected wages from our sample with those of similar graduates; we further examine how our respondents revise their expectations when provided information about actual wages.  Second, using causal mediation analysis, we test whether the consideration of a rich set of personal and professional controls, namely concerning family formation and children in addition to professional preferences, accounts for the difference in wage expectations across genders. We find that males and females overestimate their wages compared to actual ones, and that males respond in an overconfident manner to information about outside wages.  Despite the attenuation of the gender difference in wage expectations brought about by the comprehensive set of controls, gender generally retains a significant direct, unexplained effect on wage expectations.
}

\bigskip



\thispagestyle{empty}\pagebreak {\small \renewcommand{\thefootnote}{%
\arabic{footnote}} \setcounter{footnote}{0} \pagebreak %
\setcounter{footnote}{0} \pagebreak \setcounter{page}{1} }

\newpage

\setstretch{1.7} \setlength{\parskip}{6pt}

\section{Introduction}

Marked differences remain in today's labor market concerning the
professional paths of men and women. \citeasnoun{BlauKhan2017}, for example,
survey the research on the gender wage gap and find that, despite
considerable gender wage convergence during the period 1980-2010, that process was much less pronounced at the top of the wage distribution.
Further, it was the explained part of the gap that declined, indicating
gains in education and experience of women relative to men, whereas the
unexplained part of the gap did not change much.

Given the well-established and well-known gender wage gap, as well as the different paths that women and men typically follow in the labor market, rational and forward looking men and women, when asked about wage expectations about their own individual future, would generate forecasts in line with the actual gender wage gap.  The goal of our paper is, on the one hand, to examine whether this expectational wage gap is rational, in the sense of matching actual wages from comparable groups in the population, and, further, in the way that our respondents react to information about actual wages; on the other hand, we test whether the gender differences in wage forecasts of our respondents are explained by differences in their professional and personal preferences. Specifically, once we control for how individuals see themselves looking forward, not only professionally but also along personal dimensions, we want to know whether or not gender remains as a residual source of wage expectations.

Averaging the wage expectations from our respondents and contrasting them with averages of actual wages from similar population groups in the labor market, we find that both men and women overestimate wages. To gain further insight into the formation of wage expectations, we perform an experiment to examine how our respondents react to information about actual wages.  While women do not change their expectations, males react in an overconfident way to the information provided.

As commonly found in the literature, it is also the case that our male and female respondents report expecting different unconditional wages, with men having higher average wage expectations than women. Our second contribution to the literature is to examine whether this commonly found expectational wage gap vanishes once we control for a broader and more comprehensive set of controls, including detailed answers on professional and personal preferences, the latter inclusive of questions on family formation and desired number of children. Using inverse probability weighting methods in the context of causal mediation analysis to estimate the effects of gender on expected wages, we find that the broader set of covariates attenuates the direct effect of gender on wage expectations. Nonetheless, a significant and quantitatively important direct and unexplained effect of gender on wage expectations generally remains.

The paper is organized as follows. Section \ref{litsurvey} provides a literature survey on gender differences in wage expectations. Section \ref{data} introduces our data set. Section \ref{Model} outlines the methodological framework for decomposing the expectational wage gap. Results are presented in Section \ref{results} and Section \ref{conclusion} concludes.

\section{Literature Review}\label{litsurvey}

In order to simplify the exposition and facilitate the connection with our paper, we divide our overview of the literature into two strands: one that attempts to gauge the rationality/accuracy of wage expectations, and the other investigating reasons why wage expectations systematically vary with gender.

Expectations are a potentially very important input in decision making, notably when choosing schooling and investments in human capital.  As a result, some of the attempts to measure and analyze wage expectations elicited earnings expectations from students and analyzed the impact of those expectations on the individual choice of educational track going forward.

\citeasnoun{Freeman1971} and \citeasnoun{Freeman1975}, for example, showed that the expectations of students corresponded to a high degree with the performance of earlier cohorts in the labor market. This was the case both for initial wages under a variety of occupations as well as to wages after 15 years past and even at the end of the respective professional careers. He further showed that expected income differences
between occupations have an influence on the choice of education, assuming a limited set of educational alternatives.

\citeasnoun{DominitzManski1996} asked high school and college undergraduate students to complete a computerized detailed questionnaire eliciting beliefs about income levels attained under different schooling levels as well as respondents' beliefs about current earnings distributions.  They found that that students were capable of making realistic estimates of future incomes and also that  most of them expected to have positive college returns. \citeasnoun{Betts1996} similarly found that student expectations were reasonably aligned with income realizations despite considerable individual variability linked to personal traits (such as year and field of study).

Using survey data on weekly earnings and expectations from the American Survey of Economic Expectations (SEE), \citeasnoun{Dominitz1998} found that earnings expectations elicited in the form of subjective probabilities varied in sensible ways with contemporaneous earnings realizations and with other individual attributes.  Since the SEE follows earnings and earnings expectations over time, it was additionally possible to see how expectations adjusted to new income realizations.  Interestingly, respondents appeared to adjust expectations one for one when income goes down but less than one for one when they experienced an increase in income.  \citeasnoun{DasVanSoest1999} likewise examined expectations revisions over time using a Dutch panel where information on income expectations for the same household is available in consecutive years. Comparing expected and realized income changes for the same time period, they found that, on average, future income growth was significantly underestimated. In particular, people whose income decreased in the recent past tended to be too pessimistic. Negative transitory incomes were too often considered to be permanent.


A strand of research has also examined whether gender differences in wage expectations remain even after current salary information is provided.  \citeasnoun{Martin1989}, for example,
after interviewing almost 100 students from Business and Economics Departments at a mid-size American university, found that even when providing students with information about combined salaries of men and women, women expected to earn less than their male counterparts.  Here, however, wage information was provided with the survey to all the respondents; therefore, it was not possible to assess how respondents modified their wage expectations as a response to information).

Our contribution to this strand of the literature is as follows.
In line with existing work, we compare student wage expectations with realized wages for similar groups of graduates.
Further, through an information experiment, we directly analyze how information on average income for groups in the population alters wage expectations, and how such an adjustment of expectations varies by gender.
Other contributions in the literature (e.g. \citeasnoun{Dominitz1996} and \citeasnoun{DasVanSoest1999}) examine expectational adjustments under ``real world'' circumstances, thus in a situation where researchers cannot see what has effectively changed in the information set of respondents.
Since our dataset contains information pertaining to a single point in time, it is not suited for the examination of how expectations vary in response to changes in the students' own incomes.  Nonetheless, our experiment allows us to answer a more general question of how respondents react to income information on others.
Indeed, the experiment allows us to directly measure the expectational response to a change in the information set of the respondents that is fully controlled and held constant up to the new information on wage realizations. We are not aware of other experimental work aimed at examining the accuracy or rationality of adjustments in income expectations.

Since \citeasnoun{Dominitz1998}, various papers have elicited students' expectations for different population groups and tried to explain the differences. Some authors explain the differences in wage expectations and elicit wages due to differences in college and major choices.
To explain how information regarding major-specific earnings might explain educational choices, \citeasnoun{Hasting2005} evaluated a policy that provided information about college and major-specific earnings and cost outcomes in Chile. They found that information disclosure about earning outcomes changes college choice, but its impact is limited by strong non-pecuniary preferences.
Regarding gender differences, the overall view in the literature is that men and women have different own wage expectations despite having good information about earnings of their peers (\citeasnoun{SmithPowell1990}).
Despite both men and women overestimating their earnings, women's expectations seem to be more realistic when entering the labor market (\citeasnoun{HeckertEtAl2002}, \citeasnoun{Betts1996}).
Furthermore, this gender gap that closely resembles actual wage differences, prevails across subgroups, and along the entire wage distribution (\citeasnoun{Kiessling2019}).\footnote{\citeasnoun{BonnardGiret2016} find that women appear to have less accurate wage expectations concerning the long term.}

Side by side with the literature testing the rationality and accuracy of wage expectations, a different strand of work has attempted to provide reasons why expectations vary across gender. The literature has overwhelmingly confirmed that males expect higher wages than females (see e.g. \citeasnoun{BrunelloLuciforaWinterEbmen2004}).
In their pioneering work, \citeasnoun{MajorKonar1984} found that differences in career paths, pay comparisons standards, and job features explained 75\% of the variance associated to gender differences in pay expectations. \citeasnoun{Martin1989} and \citeasnoun{McFarlin1989} found a similar gender disparity in expected entry-level wages among senior students in business-related fields. As stated earlier, an expectational wage gap may simply be a reflection of rational forecasts of an actual wage gap, the latter very well documented as well in the empirical literature.  Nonetheless, and in view of the role expectations take in informing decisions, establishing the causes of the expectational wage gap by gender remains an important open question.

Gender differences in wage expectations have not been limited to the American context. \citeasnoun{Menon2012} provided evidence of gender differences in employment and wage expectations in Cyprus. They showed that female students have lower earnings expectations than their male counterparts, and that these expectations were realistic even among Cypriot educated women. Using a survey conducted at a major Italian university, \citeasnoun{FilipinIchino2005} show that expected wages are consistent with realized gender wage gaps at the entry level, but students underestimate the wage gap along the career.
Moreover, they found that females always expect a higher wage gap than males, and a larger fraction of females attributes the causes of this gap to employers' discriminatory preferences.
In a recent study using a sample of over 15,000 German students, \citeasnoun{Kiessling2019} documented a large gender gap in wage expectations that resembles actual wage differences. Moreover, this gap in wage expectations remains across population subgroups and along the wage distribution.
While gender differences in wage expectations are present at the start of people's career, they increase over the prospective life cycle.
\citeasnoun{HogueDuBoisFox2010} argue that salary expectations start even before an individual begins to work. Using a sample of about 500 undergrad students in the US, they show that women expected to be paid less than men at the beginning and it increases at the peak of their careers.

While these gender differences in wage expectations have been documented extensively in the literature, the key question is indeed how gender differences in wage expectations are explained. Multiple studies provide different answers.
We discuss next five of the main causes that might drive gender differences in wage expectations, and determine how family and career expectations contribute to explain this gap.

First, individual characteristics as well as job preferences may lead men and women to chose different jobs and make different career decisions. For example, in \citeasnoun{HogueDuBoisFox2010}, men and women are conscious of the pay implications that choosing a female- or a male-dominated job will imply. They show that job intentions (partially) account for  the gender differences in pay expectations. Individuals who intended to hold female-dominated jobs expected to be paid less than those pursuing male-dominated jobs.
Along similar lines, \citeasnoun{GasserFlintTan2000} found that men have higher promotion expectations for male- and neutral-oriented jobs than their female counterparts.
Existing studies have not been limited to female-male job classifications, however.
\citeasnoun{Osikominu2018} confirm that women have consistent lower wage expectations than men across different education programs such as STEM and non-STEM fields. Nonetheless, they found that differences in wage expectations were not explained by the probability of choosing a STEM major.

Second, psychological features such as self-perceptions, self-esteem and self-efficacy have also been examined.
The psychological literature, in particular, is very rich regarding self-related theories and has shown that wage expectations and self-views are correlated with job attributes and pay expectations (\citeasnoun{HogueDuBoisFox2010}).
These authors attribute lower female self-wage expectations to the fact that women see themselves performing a female-dominated job, or to their belief that they are less competent at work than men.
They also suggest that women believe holding a male-dominated job creates a pay advantage, and in reality, holding a male-dominated job boosts women's pay.
Contrasting with these findings, \citeasnoun{ZhangZheng2019} found that gender differences in self-views favors females in China. They attribute this evidence to their sample -- in which most women hold a career track position and have longer average years of education than the national average in China.

A third explanation for gender differences in wage expectations is rooted in gender differences in attitudes towards preference for competition and negotiation skills.
Using a sample of about 1500 Swiss lower-secondary school students, \citeasnoun{Buser2017} found that, at all levels of the ability distribution, willingness to compete is associated with choosing more challenging options which, in turn, leads to higher-paid careers.
\citeasnoun{Kleinjans2009} pointed out that female's distaste for competition decreases educational achievement and partially explains the gender segregation in occupational fields, particularly in Law, Business \& Management, and Health and Education.
Using tournament data, \citeasnoun{Niederle2007} showed that women dislike competition and that men are substantially more overconfident than women.  This behavior towards competition may carry over to other fields and situations, for example to salary negotiations and negotiation skills and strategies.
\citeasnoun{Barron2003} showed that differences across male and female's salary requests and in salary negotiation behavior are explained by beliefs about self, entitlement, other opportunities, and one's value in the workplace.

A fourth explanation concerns gender differences in attitudes related to work as well as family values and aspirations.
Individuals might be already aware at an early age of their career and family plans, and therefore internalize those in their future decisions in the labor market.
As a result, perhaps expecting to work fewer hours or taking jobs that help reconcile family and work (\citeasnoun{Goldin2014}), women may have lower salary expectations, and lower wages compared to men (\citeasnoun{Adda2017}).
\citeasnoun{Chevalier2007} documented that child rearing and career break expectations accounted for about 10\% of the gender wage gap in the UK.
For these authors, because women have child rearing preferences and expect to take a career break, they reduce their search and expect a lower wage. As consequence, they are more likely to be employed in a poorer job, less willing to move across jobs, and more likely to obtain a lower wage.
\citeasnoun{Heckert2002} also pointed out that wage expectations are affected by time preferences for childrearing but also for conciliating weekly hours worked with family responsibilities.
Using very detailed information on career plans and earning expectations of college business school seniors, \citeasnoun{BlauFerber1991} provide evidence women expect to work fewer years than men. However, gender differences in expected earnings have any effect on the number of years that women expect to work in the labor market.
In our paper, we paid particular attention to these factors and investigate how much family and career aspirations explain of the gender differences in wage expectations.

A fifth explanation examines factors such as career referents, stereotypes, social comparisons, and perceived discrimination or perceptions about pay standards.
\citeasnoun{GibsonLawrence2010} found that women have lower expectations than men even when they identified high-level referents and even when those referents are women.
Related to this, \citeasnoun{McFarlin1989} show evidence that same-sex comparisons are a stronger predictor of career-entry pay expectations than opposite-sex comparisons.
\citeasnoun{JacksonGardnerSullivan1992} found that fair pay standards was the main predictor for explaining differences in wage expectations.
\citeasnoun{Orazem2003} show that lower reservation wages for women are explained because women perceived lower employment opportunities than men which are due to perceived discrimination in the work force.
According to these authors, the adverse effects of perceived discrimination might be minimized as women gather more information about labor market prospects, so they can increase their reservation wages and actual wages.
In this line, also a recent laboratory experiment of \citeasnoun{BordaloEtAl2019} showed that stereotypes affect self-beliefs and own performance.
It is therefore important to know whether individuals have accurate wage perceptions.

Our contribution to this strand of the literature is as follows.
First, we provide a comprehensive overview and compare the role of career and family aspirations in determining wage expectations. As detailed before, previous studies have pointed out the importance of these factors when studying their role individually but were not able to disentangle their individual contributions once jointly considered.
Second, we compare differences in expectations with real wage differences, even when information is provided.
In our new survey among students of two Swiss institutions of higher education, we cover not only professional aspirations and preferences but also personal ones. We included detailed questions about the respondent's intent in forming a family, number of children s/he would like to have, degree of labor market attachment in the presence of children, among others, which allow us to determine exactly the importance of these covariates for eliciting wage expectations.
More precisely, we contribute directly to the literature addressing in the fourth explanation above although we are also part of the contributions that focus on individual characteristics and job preferences, the first explanation.
Further, to provide more sophisticated empirical methods than those conventionally used in decomposing the (expectational) gender wage gap. Specifically, by resorting to casual mediation analysis, we are better able to address endogeneity issues as well as to allow for nonlinear relations between our covariates and confounders and wage expectations.
		
\section{Data}\label{data}

Our data was collected by running a detailed survey among undergraduate
students in two Swiss institutions of higher education, namely the
Business School of the Bern University of Applied Sciences (BUAS), and the
Faculty of Economic and Social Sciences of the University of Fribourg. The answers were
collected on paper. Individual classes were visited and the survey was
presented to students for immediate completion and collection. Data
collection took place mostly during the first week of the Spring semester of
2017 in order to maximize the response rate. All undergraduate classes of the
Business Administration (BBA) and Business and IT (BWI) degrees were visited
at BUAS; at the University of Fribourg, respondents attended one of two
large introductory statistics classes in the Economics program. Respondents
in Fribourg were enrolled in three different degree programs: Bachelor in
Economics (VWL), Business Administration (BBA) and Communications (KOMM).

The questionnaires comprised three groups of questions. Students were first
asked about general information, such as age, gender, nationality, degree
chosen and whether they were enrolled in a part- or full-time program. This was followed by two separate blocks of questions about professional and
personal matters. The professional section asked students about their
preferences regarding a variety of job attributes, whether they intended to
work part- or full-time upon completion of their degree, industry and
occupation where they expect to be working, expected wage upon completion of
the degree and three years after graduation, if they would rather hold a
management or a consulting/supporting position. The personal section
inquired about the intent of forming a family, number of desired children
and associated intended degree of labor market participation in the presence
of children, the labor market attachment of both parents during different
stages of childhood (daycare, Kindergarten and primary school ages of the
respondent), type of residence, family composition, and educational
attainment of both parents, among other questions. In total, our sample comprises 865 observations from both educational
institutions combined.

While all questionnaires had these three groups of questions, the order in
which professional and personal questions appeared was randomized. In the
control version of the questionnaire, general questions were followed by
professional questions, with the personal block appearing at the end. In
another version of the questionnaire, the second group of questions was
about personal matters and the professional block only showed at the end. We
labeled this version as \textquotedblleft the different
order\textquotedblright\ version. A third version of the questionnaire
retained the question order of the control version but introduced a bar
graph with information on monthly gross income in the private sector, according to age and gender.
This is the \textquotedblleft information\textquotedblright\ version of the
survey. It is important to see that this information was not necessarily helpful for forming expectations about the own wage directly upon finishing university of three years thereafter. This because it neither focused on university graduates nor on years in the labor market.
The information version of questionnaire can be found in Appendix A.\footnote{As indicated above, the control version was obtained by eliminating the bar graph from page 1.}

\section{Methodological framework}\label{Model}

The decomposition of wage gaps across gender aims at disentangling the total gap into an explained component that can be attributed to differences in observed labor market relevant characteristics and an unexplained remainder. In addition to the classical linear decomposition of \citeasnoun{Blinder73} and \citeasnoun{Oaxaca73},  non-parametric decomposition methods have for instance been proposed by \citeasnoun{DinardoFortinLemieux96}, \citeasnoun{BarskyBoundCharlesLupton01}, \citeasnoun{Fr07}, \citeasnoun{Mora2008}, and \citeasnoun{Nopo2008}, as well as methods for decompositions at quantiles rather than means, see \citeasnoun{JuhnMurphyPierce93}, \citeasnoun{DinardoFortinLemieux96}, \citeasnoun{MachadoMata05}, \citeasnoun{Melly05}, \citeasnoun{FirpoFortinLemieux07}, \citeasnoun{ChernozhukovFernandezValMelly2009}, and \citeasnoun{FirpoFortinLemieux09}. However, such progress in estimation methods stands in contrast to the widespread ignorance of identification issues, in particular the endogeneity of observed characteristics, as for instance pointed out in \citeasnoun{FortinFortinLemieux2011}.

Following \citeasnoun{huber15} and \citeasnoun{HuberSolovyeva2018b}, we formulate the decomposition in the context of a causal model for mediation analysis which allows explicating endogeneity issues. Mediation analysis, as for instance discussed in \citeasnoun{BaKe86}, aims at disentangling the causal mechanisms through which an explanatory variable affects an outcome, with mediators being intermediate outcomes lying on the causal pathway between the explanatory variable and the outcome. Applied to decompositions, gender is the explanatory variable at the beginning of any individual's causal chain affecting expectational wage, because it is determined at or prior to birth. Choice of study program, career preferences, and family plans, on the other hand, are mediators (often referred to as observed characteristics in the wage decomposition literature), because they occur later in life and are thus potentially influenced by gender, while the mediators themselves likely affect wage expectations. Given this causal structure, the explained component in the decomposition literature corresponds to the indirect effect of gender on wage expectations that operates through these mediators. Conversely, the unexplained component equals the direct effect of gender on wage expectations that operates through unobserved mediators like unmeasured personality traits.

\begin{figure}[!htp]
\centering \caption{\label{fig1med}  Graphical representation of the decomposition}
\begin{center}
{\small \includegraphics[scale=1, trim={4cm 23cm 8cm 4cm}]{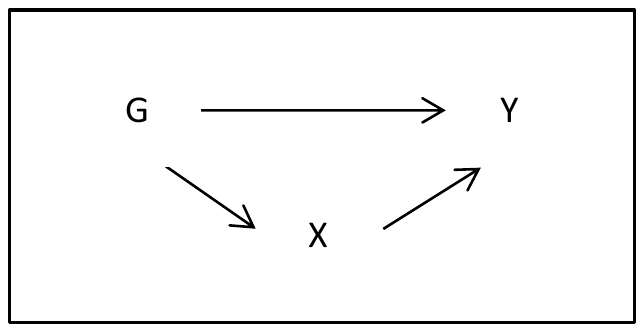}}
\end{center}
\end{figure}

More formally, let $G$ denote a binary dummy for gender, $Y$ the wage expectations outcome and $X$ a vector of observed mediators. $G$ may affect $Y$ indirectly via its effect on $X$, i.e.\ by a causal mechanism related to observed characteristics. For instance, gender might have an effect on expected wages because females and males target different job types. $G$ might influence $Y$ also directly, i.e.\ through factors not observed in $X$.  A graphical representation of this causal framework is given in Figure \ref{fig1med}, with arrows representing causal effects: $G$ influences $Y$ either through $X$ or directly. For defining the effects of interest, we denote by $Y(g)$ and $X(g)$ the potential outcomes and mediators when exogenously setting gender $G$ to value $g$, with $g$ $\in$ $\{1,0\}$.\footnote{See  for instance \citeasnoun{Rubin74} for an introduction to the potential outcome framework.} Then, $E(X(1))-E(X(0))$ gives the average causal effect of $G$ on mediators $X$, while $E(Y(1))-E(Y(0))$, corresponds to the (total) average causal effect of $G$ on $Y$, represented by the sum of direct and indirect (i.e.\ operating through $X$) effects.

As in \citeasnoun{RoGr92} and \citeasnoun{Pearl01} (among many others), we further refine the potential outcome notation to adapt it to the mediation framework: Let $Y(g)=Y(g,X(g))$, which explicates that the potential outcome is influenced by the group variable both directly and indirectly via $X(g)$. We can thus express the total effect of $G$ on $Y$ as $E(Y(1))-E(Y(0))=E[Y(1,X(1))]-E[Y(0,X(0))]$ in order to decompose the latter into direct and indirect effects. That is, the difference in potential outcomes due to altering $X(1)$ to $X(0)$ while keeping gender fixed at $G=1$ yields the indirect effect (denoted by $\psi$),  while modifying gender $G$ and keeping the characteristics constant at $X(0)$ yields the direct effect ($\eta$). Summing both up gives the total causal effect, as formally expressed below:
\begin{eqnarray}
E[Y(1,X(1))]-E[Y(0,X(0))]  =  \underbrace{E[Y(1,X(1))]-E[Y(1,X(0))]}_{\psi} \nonumber \\
						  	+  \underbrace{E[Y(1,X(0))]-E[Y(0,X(0))]}_{\eta}
\end{eqnarray}
Depending on whether (the wage expectations of) males or females are considered as reference group with $G=1$, the magnitudes of $\psi$ and $\eta$ may differ due to interaction effects between $G$ and $X$, i.e.\ effect heterogeneities in gender. In the application provided in Section \ref{results}, we present the results when considering both females and males as reference group.

The Oaxaca-Blinder decomposition consistently estimates the indirect and direct effects as the explained and unexplained components, respectively, when both $G$ and $X$ are exogenous. This rules out confounding of the gender-outcome, gender-mediators, or mediators-outcome relationship. In addition, the decomposition requires the wage expectations to be linear in the mediators is also required. Assumption 1 formalizes these restrictions.\vspace{5pt}\\
\textbf{Assumption 1 (sequential independence):}\newline
(a) $\{Y(g',x), X(g)\}  \bot G $ for all $g',g \in \{0,1\}$ and $x$ in the support of $X$,\\
(b) $ Y(g',x) \bot  X | G=g  $ for all $g',g \in \{0,1\}$ and $x$ in the support of $X$,\\
(c) $Y(g,X)$ is linear $X$ for $g \in \{0,1\}$,
\vspace{15pt}
\newline
where `$\bot$' denotes statistical independence. Under Assumption 1(a), $G$ quasi-random, i.e.\ there are no variables affecting both $G$ on the one hand and $Y$ and/or $X$ on the other hand. Under Assumption 1(b), observed characteristics like education quasi-random within gender, i.e.\ given $G$, so that there are no affecting both $X$ and $Y$.\footnote{Assumptions 1(a) and 1(b) could be relaxed to mean independence when considering average wage gaps, while full independence is required for decompositions of quantiles.} Assumption 1(c) imposes potential outcomes to be linear in $X$.

As also discussed in \citeasnoun{HuberSolovyeva2018b},  under Assumption 1(a), $E(X(g))=E(X|G=g)$, while under Assumptions 1(a), 1(b), and 1(c),  $E[Y(g,x)]=E(Y|G=g,X=x)=c_g+x\beta_g$, with $c_g$ denoting a gender-specific constant and $\beta_g$ a vector of gender-specific coefficients on $X$. By iterated expectations, $E[Y(g,X(g'))]=c_g+E(X|G=g')\beta_g$ for $g,g'$ $\in$ $\{0,1\}$. Therefore,
\begin{eqnarray}
\psi&=&E[Y(1,X(1))]-E[Y(1,X(0))]=[E(X|G=1)-E(X|G=0)]\beta_1, \label{ob1}\\
\eta&=&E[Y(1,X(0))]-E[Y(0,X(0))]=c_1-c_0+E(X|G=0)(\beta_1-\beta_0).\label{ob2}
\end{eqnarray}
The probability limits of the explained and unexplained components in the linear Oaxaca-Blinder decomposition correspond to the left hand expressions in \eqref{ob1} and \eqref{ob2}, respectively.

In this paper, we also consider a semiparametric propensity score weighting approach for causal mediation analysis suggested in \citeasnoun{Huber2014}, that improves on the Oaxaca-Blinder decomposition in two dimensions. First, it does not impose linearity of the outcome in the mediators and second, it allows controlling for observed confounders not influenced by the treatment, henceforth denoted by $W$. We therefore control for a range of socio-economic variables (including age, nationality, number of siblings,  parental education and occupation, material wellbeing) to at least mitigate the endogeneity of gender (which is not necessarily random in the two institutions considered) and the mediators. Among $W$ we also include the randomized version of the questionnaire, i.e.\ indicators for whether the order of professional and personal questions was reversed or whether a graph with information on median monthly gross earnings by age and gender were shown. While the questionnaire version is not related to $G$ due to randomization, it might potentially affects both $X$ and $Y$.

Formally, out estimation approach is consistent under Assumption 2, which has among others also been considered in \citeasnoun{ImKeYa10}.\vspace{5pt}\\
\noindent\textbf{Assumption 2 (sequential conditional independence):}\newline
(a) $\{Y(g',x), X(g)\}  \bot G | W$ for all $g',g \in \{0,1\}$ and $x$ in the support of $X$,\\
(b) $ Y(g',x) \bot  X | G=g, W=w  $ for all $g',g \in \{0,1\}$ and $x,w$ in the support of $X,W$,\\
(c) $\Pr(G=1| X=x, W=w)>0$ and $0<\Pr(G=1| W=w)<1$ for all $x,w$ in the support of $X,W$.
\vspace{15pt}\newline
Assumptions 2(a) and (b) require that conditional on $W$, no unobserved confounders either jointly affect $G$ and $Y$, $G$ and $X$, or $X$ and $Y$ given $G$. We acknowledge that this may not hold in our empirical application presented further below, given the limited number of observed control variables. Yet, including control variables likely improves upon conventional wage decompositions that do not account for any form of confounding. Assumption 2(c) is a common support condition, requiring that the conditional probability of belonging to the reference group ($G=1$) given $X,W$ is larger than zero, while the conditional probability given $W$ must neither be zero nor one. The latter restriction means that for each value of $W$, there exist both females and males in the population.

Under Assumption 2, it follows from the results on identification of direct and indirect effects by inverse probability weighting (IPW) in \citeasnoun{Huber2014} that
\begin{eqnarray}
\psi&=&E\left[\frac{Y\cdot G}{\Pr (G=1|W)} \right]-E\left[\frac{Y\cdot G}{\Pr (G=1|X,W)} \cdot \frac{1-\Pr(G=1|X,W)}{1-\Pr(G=1|W)}\right],\label{ipw.ex}\\
\eta&=&E\left[\frac{Y\cdot G}{\Pr(G=1|X,W)} \cdot \frac{1-\Pr(G=1|X,W)}{1-\Pr(G=1|W)}\right]-E\left[\frac{Y\cdot (1-G)}{1-\Pr (G=1|W)} \right].\label{ipw.un}
\end{eqnarray}
In our application, we estimate $\Pr(G=1| X=x, W=w)$ and $\Pr(G=1| W=w)$ by probit regressions and $\psi$ and $\eta$ by normalized sample analogues of \ref{ipw.ex} and \ref{ipw.un}, respectively, implying that the weights of observations within treatment states sum up to one in our sample. Furthermore, we drop observations with estimated propensity scores below 2\% (or 0.02) and above 98\% (or 0.98) to prevent that some observations receive too extreme weights in the estimation of direct and indirect effects.

In addition to the expectational wage gap decomposition by gender, the empirical analysis investigates whether the randomized questionnaire version affects the wage expectations of females and males differently. The questionnaire version is therefore once regarded as control variable for mediator-outcome confounding in the decomposition, and once as treatment variable to assess its causal effect on wage expectations.

\section{Results}\label{results}

We begin our empirical analysis by comparing average wages in our sample to averages of realized wages from comparable samples, as follows.  The first two rows of Table \ref{descript2} contain average wages from the FH-Lohn, a yearly survey of wages of alumni of Swiss Universities of Applied Sciences. Survey participation is voluntary and elicited by email, and further completed in an electronic format.  In Table \ref{descript2} below (3$^{rd}$ and 4$^{th}$ rows), we include the average of wages reported by alumni in the surveys of 2016 and 2017. For 2016, we only had data on average wages for all former students that answered the survey, thus comprising all ages and experience. For 2017, it was possible to confine the search to alumni younger than 30 years of age, reaching therefore an age range and presumed experience closer to that of our student respondents. It was further possible to separate the average wages by gender and degree, and the numbers reported correspond to Business School former students, specifically to alumni who graduated in Business Administration.

\begin{table}[!htbp]
	\caption{\label{descript2} Descriptives: Realized Wages and Wage Expectations} {\footnotesize
		\begin{center}
			\begin{tabular}{r c c c c c c c}
				\hline \\
				Means			& \multicolumn{1}{c}{Sample} & \multicolumn{1}{c }{Years of}   & \multicolumn{1}{c}{Obs} & \multicolumn{1}{c}{Male} 	  & \multicolumn{1}{c}{Obs} & \multicolumn{1}{c}{Females} & \multicolumn{1}{c }{\% Change} \\
				& \multicolumn{1}{c}{source} & \multicolumn{1}{c }{Experience} &     &	  &  &    & \multicolumn{1}{c}{Males-Females} \\
				\hline\\
				
			    Realized					& FH-Lohn (2016)		    &	all						 & 	 26 & 6116.85 & 18     & 5667.54 & 7.93\% \\
				Wages 						& FH-Lohn (2017)			&	less than 30 years old   & 111     & 6099.54 &100      & 5776.46 & 5.59\% \\
				\cmidrule{2-8}
				Wage 			 			& Fribourg University  	 	& 0							 & 497	& 6858.15 &   360  & 6252.43 & 9.69 \% \\
				Expectations     			&  \& BFH 				  	& 3							 & 498	& 8418.42 &   360  & 7546.18 & 11.56 \%  \\
				\cmidrule{2-8}
				(own survey)	  			& Only BFH				 	& 0 						 & 391	& 6890.35 &   249  & 6424.20 & 7.26 \% \\
											& 						  	& 3							 & 392	& 8430.80 &   249  & 7564.76 & 11.45 \% \\
				\cmidrule{2-8}
											& \multicolumn{7}{c}{Information Treatment (BFH only)}						  \\
				\cmidrule{2-8}
											& Exposed to information    & 0 						 &	130	& 6947.12 & 		75		& 6291.67 & 10.42 \% \\
											&	(treated group)			& 3							 &	129	& 8551.36 & 		75		& 7553.33 & 13.21 \% \\
											& No exposed to information & 0 						 &	137	& 6889.60 & 		79		& 6465.19 & 6.56 \% \\
											&	(control group)			& 3							 &	140	& 8333.04 & 		80		& 7518.75 & 10.83 \% \\
				\hline
			\end{tabular}
	
		\end{center}
		\par
		Note: This table uses information from the survey collected by the Association of Universities of Applied Sciences (FH-Lohn), and our own survey.}
\end{table}

Rows 3 through 6 include average wages from our own sample, first from the full sample of BFH and the University of Fribourg combined, and then for BFH alone.  The bottom 4 rows partition the BFH into a subsample from the information treated group (which received information on outside wages in their questionnaire) and the control subsample (devoid of outside wage information).

Our sample confirms one stylized fact from the literature, namely the existence of an expectational gender wage gap (rows 3 through 10, last column).  In particular, for our overall sample, the expectational gender wage gap is 9.7\%,  concerning expected wages upon graduation, and 11.6\%, for expected wages three years thereafter.  (Actual wages from FH-Lohn are also in line with the well-established (raw) gender wage gap as can be seen in rows 1 and 2, last column.) Expected wages increase over time, with expectations 3 years ahead of graduation systematically exceeding those for graduation wages (rows 3 through 10, along ``Male'' and ``Female'' columns). We cannot really compare the 2016 and 2017 sample averages of FH-Lohn because they relate to two different groups: the 2016 average wage was an average over all graduates who took the survey whereas the 2017 figure is for graduates under 30.  It is somewhat surprising that the 2016 average wage, excluding more experienced individuals, amounts to a higher average wage than the 2017 average wage, pertaining to graduates of all ages and experience. This may simply reflect greater sample attrition for higher age groups.

For the purpose of comparing realized wages and expectations,  the two most similar groups in Table \ref{descript2} are 2017 average realized wages from FH-Lohn  (younger than 30), and the BFH student respondents. Table \ref{descript3} computes the percentage excess of the average wage expectations from different BFH subsamples relative to the 2017 average wage from FH-Lohn, the latter from row 2 in the preceeding Table.

\begin{table}[!htbp]
	\caption{\label{descript3} Deviations from Reality: Percentage differences between Expected and Realized Wages} {\footnotesize
		\begin{center}
			\begin{tabular}{rccc}
				
				\multicolumn{3}{c}{Wage differences ($w^e$ - $w$)/$w$, \%} \\
				
				\hline \\
				
				&  Men      & Women   \\			
				\hline	\\									      								
				&  \multicolumn{2}{c}{0 years of experience} \\
				\cmidrule{2-3}									
				Only BFH										  & 12.44    & 8.24   \\
				Exposed to information (treated group)			  & 13.09    & 8.92    \\
				No exposed to information (control group)		  & 12.95    & 11.92   \\
				\cmidrule{2-3}								
				&  \multicolumn{2}{c}{3 years of experience} \\
				\cmidrule{2-3}											
				Only BFH										  & 38.02    & 30.64  \\
				Exposed to information (treated group)			  & 40.2    & 30.76   \\
				No exposed to information (control group)		  & 36.62    & 30.16  \\
				\hline
			\end{tabular}
		\end{center}
		\par
		Note: Uses information from Table \ref{descript2} to compute percentage differences.}
\end{table}

As the numbers in Table \ref{descript3} indicate (all entries are positive), both men and women overestimate their wages relative to those of comparable graduates.  Comparing expectations and actual wages for workers without experience, for example, the average of BFH wage expectations for males in the full sample exceeds by 12.44\%  actual earnings of similar graduates in 2017. For females, this gap amounts to 8.24\%.  The gap widens once we consider wages with three years' experience.  For the whole of BFH, average expected wages exceed realizations by 38\% for males, and by 30.6\% for females. Thus, both males and females are overconfident in that expectations systematically exceed realizations from comparable groups.

Table \ref{descript3} also shows how average wage expectations changed for the group that received outside wage information.  For treated males, average expected wages appear to increase, and this is true for no experience or three years on.  For females, the change in average expected wages is more subtle and there is even a decline for the shorter time horizon.  We next proceed to a rigorous analysis of the causal effects of the questionnaire randomization, described earlier.

\begin{table}[!htbp]
	\caption{\label{descripttreat}  Descriptives for experiment} {\footnotesize
		\begin{center}
			\begin{tabular}{r|c|ccc|ccc|c}
				\hline \hline
				& control & \multicolumn{3}{c|}{ treatment: information}  & \multicolumn{3}{c}{treatment: order} & \\
				& mean  & mean  & dif & pval & mean  & dif & pval & missing \\
				\hline
female & 0.39 & 0.41 & 0.02 & 0.59 & 0.47 & 0.08 & 0.06 & 1  \\
  age & 23.06 & 23.25 & 0.19 & 0.35 & 23.25 & 0.20 & 0.34 & 0  \\
  Swiss & 0.89 & 0.86 & -0.03 & 0.23 & 0.89 & 0.00 & 0.85 & 0 \\
  has siblings & 0.89 & 0.92 & 0.03 & 0.20 & 0.95 & 0.06 & 0.01 & 5  \\
  mom has higher education & 0.24 & 0.20 & -0.04 & 0.22 & 0.20 & -0.04 & 0.26 & 8  \\
  dad has higher education & 0.39 & 0.40 & 0.01 & 0.80 & 0.37 & -0.02 & 0.63 & 13  \\
  mum worked full time when I was 4-6 & 0.15 & 0.16 & 0.01 & 0.75 & 0.14 & -0.01 & 0.68 & 7  \\
  mum worked part time when I was 4-6 & 0.44 & 0.45 & 0.01 & 0.80 & 0.41 & -0.03 & 0.42 & 7  \\
  wellbeing & 2.41 & 2.30 & -0.11 & 0.08 & 2.35 & -0.05 & 0.35 & 5  \\
  home owner & 0.45 & 0.40 & -0.06 & 0.18 & 0.45 & -0.01 & 0.89 & 8  \\
  program: business admin & 0.72 & 0.71 & -0.01 & 0.82 & 0.72 & -0.00 & 0.90 & 2  \\
  program: economics & 0.04 & 0.02 & -0.02 & 0.28 & 0.02 & -0.01 & 0.36 & 2  \\
  program: communication & 0.06 & 0.08 & 0.02 & 0.29 & 0.07 & 0.01 & 0.70 & 2  \\
  program: business IT & 0.15 & 0.13 & -0.02 & 0.41 & 0.14 & -0.01 & 0.70 & 2  \\
				\hline
                 number of observations & 298 & \multicolumn{3}{c|}{ 277} & \multicolumn{3}{c|}{ 293}&\\
			\end{tabular}
		\end{center}
		\par
		Note: `mean', `dif', and `pval' reports the respective means, mean differences and p-values of the mean differences. `missing' provides the number of missing observations in the respective variable.}
\end{table}

In what follows, we have divided the variables of our dataset into the following categories: control variables $W$, mediators $X$, outcomes $Y$ (the expected gross wage category directly after the studies or 3 years later), and $G$ for gender.

To investigate whether randomization was successful, Table \ref{descripttreat} reports the means of the control variables $W$ separately for the control group, for questionnaires with the information treatment, and for those with a reversed order of personal and professional questions. Mean differences between the means of the respective treatment group and control group as well as p-values of mean difference tests are also provided. Balance appears to be decent (albeit not perfect), as only 3 differences are significant at the 10\% level and only 1 is at the 1\% level.

\begin{table}[!htbp]
	\caption{\label{intereffects}  Intervention effects} {\footnotesize
		\begin{center}
			\begin{tabular}{r|ccc|ccc|ccc|ccc}
				\hline \hline
				& \multicolumn{6}{c|}{ treatment: information} & \multicolumn{6}{c}{treatment: order} \\
				& \multicolumn{3}{c|}{ female} & \multicolumn{3}{c|}{male} & \multicolumn{3}{c|}{ female} & \multicolumn{3}{c}{male} \\
				& est & se & pval & est & se & pval & est & se & pval & est & se & pval \\
				\hline
				&  \multicolumn{12}{c}{outcome: gross wage category after studying}\\
				\hline
				mean differences & -0.19 & 0.33 & 0.57  & 0.23 & 0.27 &   0.38 & 0.26 & 0.28 & 0.37 & -0.09 & 0.27 & 0.74 \\
				OLS with controls & -0.26 & 0.33 & 0.43 & 0.27 & 0.29 & 0.36 &  0.22 & 0.28 & 0.42 &  -0.18 &  0.28 & 0.53 \\
				double lasso              & -0.36 & 0.33 &  0.28 & 0.23 & 0.30 & 0.44 & 0.18 & 0.28 & 0.53 & -0.14 &  0.28 &  0.63 \\
				mean among controls & \multicolumn{3}{c|}{5.97} & \multicolumn{3}{c|}{ 7.11} & \multicolumn{3}{c|}{5.97} & \multicolumn{3}{c}{7.11}\\
				\hline
				&  \multicolumn{12}{c}{outcome: gross wage category 3 yrs after studying}\\
				\hline
				mean differences &  0.14 & 0.33 &  0.68 & 0.62 & 0.32 & 0.05 & 0.14 & 0.32 & 0.66  & -0.21 & 0.31 & 0.51  \\
				OLS with controls & -0.01 &  0.35 & 0.97 & 0.61 & 0.34 & 0.08 & 0.05 &  0.32 & 0.86 &  -0.18 &   0.32 &   0.58 \\
				double lasso &  -0.07 &  0.34 &  0.84 & 0.60 & 0.34 & 0.07 & 0.05 & 0.31 & 0.88 & -0.19 &  0.32 & 0.54 \\
				mean among controls & \multicolumn{3}{c|}{8.47} & \multicolumn{3}{c|}{9.91} & \multicolumn{3}{c|}{8.47} & \multicolumn{3}{c}{9.91}\\
				\hline
			\end{tabular}
		\end{center}
		\par
		Note: `est', `se', and `pval' reports the ATE estimates, heteroscedasticity robust standard errors, and p-values for the mean difference estimator (`mean differences'), OLS controlling for $W$ (`OLS with controls'), and doubly robust estimation based on separate lasso estimations of the propensity score and the conditional mean outcome (`double lasso'). The mean value of the outcome in the control group (`mean among controls') is also reported.}
\end{table}

Table \ref{intereffects} reports the results of the treatments on either outcome separately for females and males. The first row (`mean differences') presents the experimental estimate based on mean differences in outcomes between the respective treatment group and the control observations.\footnote{Wage expectations after studying and three years later are not reported for 10 and 9 observations, respectively, that are dropped from the analysis.} The second row (`OLS with controls') provides the estimated when linearly conditioning on $W$ based on OLS to control for any imbalances in the potential confounders.\footnote{38 observations with either missings in $W$ or $Y$ are dropped from the analysis.} The third row (`double lasso')presents the results when using (double) lasso-based estimation of the treatment propensity score  $\Pr(G=1|W)$ and of the outcome $E(Y|G,W)$ to estimate the treatment effect by semiparametric doubly robust estimation, see \citeasnoun{Bellonietal2017}. We to this end use the `rlassoATE' command with its default options of the `hdm' package by \citeasnoun{ChernozhukovHansenSpindler2016} for the statistical software `R'. This method controls for elements in $W$ in a data-driven way under the assumption of approximate sparsity, i.e.\ that relatively few variables suffice for tackling most of treatment-outcome confounding.

Concerning the information treatment, it is worth noting that the gender wage gap of the displayed age categories 20-29 and 30-39 is roughly in line with the average expectational wage gap in the group with the control version of the experiment (see `mean among controls' for males vs.\ females). However, the expected average wage levels in the control groups are considerably lower than in the graph of the information treatment. The effect estimates (`est') suggest that the information treatment increased males' wage expectations three years after studying by roughly 0.6 categories (or 300 CHF). These estimates are marginally statistically significant at the 10\% level across the three methods considered, see the heteroscedasticity robust standard errors (`se') and p-values (`pval'). The information treatment therefore appears to increase the expectational wage gap between males and females. Furthermore, it exacerbates over-confidence among males, as early career wages among university graduates are actually lower than expected by both males and females (even without information treatment), see Table \ref{descript3}. 
In contrast, the information treatment did neither statistically significantly affect males' wage expectations directly after studying (albeit point estimates are again positive)  nor female expectations in either period. Secondly, reversing the order of the professional and personal questions did not show any significant impact on wage expectations.

Finally, we address our second question, namely whether or not the inclusion of a broad set of controls, focusing not only on professional preferences but also on personal ones, suffices to account for the direct, unexplained effect of gender on wage expectations.

Table \ref{descript1} reports descriptive statistics for control variables $W$, mediators $X$, and  outcomes $Y$, separately by gender $G$. The first 4 columns provide the means of (non-missing values of) the respective variables by gender, mean differences across gender, and the p-values of differences-in-means tests for the original sample. We observe that females and males differ importantly in a range of characteristics like the choice of study program, age, targeted industry and occupation, as well as job expectations and job views.

\begin{table}[!htbp]
	\caption{\label{descript1}  Descriptives and balance tests for covariates and mediators} {\scriptsize
		\begin{center}
			\begin{tabular}{r|cccc|cc}
				\hline \hline
				& \multicolumn{4}{c|}{Original sample} & \multicolumn{2}{c}{After re-weighting} \\
				& mean females & mean males & difference  & p-value & difference  & p-value \\
				\hline
				\multicolumn{7}{c}{Control variables $W$}\\
				\hline
				age & 22.84 & 23.44 & 0.60 & 0.00 & -0.01 & 0.92 \\
				Swiss & 0.88 & 0.88 & -0.00 & 0.92 & -0.01 & 0.62 \\
				has siblings & 0.92 & 0.92 & -0.00 & 0.85 & -0.03 & 0.14 \\
				mom has higher education & 0.23 & 0.20 & -0.03 & 0.31 & 0.00 & 0.93 \\
				dad has higher education & 0.42 & 0.37 & -0.05 & 0.13 & -0.00 & 0.98 \\
				mum worked full time when I was 4-6 & 0.21 & 0.11 & -0.09 & 0.00 & 0.01 & 0.57 \\
				mum worked part time when I was 4-6 & 0.44 & 0.43 & -0.01 & 0.71 & -0.00 & 0.93 \\
				wellbeing & 2.35 & 2.36 & 0.01 & 0.80 & 0.04 & 0.39 \\
				home owner & 0.44 & 0.43 & -0.01 & 0.83 & -0.04 & 0.11 \\
				treatment: information & 0.31 & 0.33 & 0.01 & 0.67 & 0.02 & 0.41 \\
				treatment: order & 0.37 & 0.31 & -0.06 & 0.07 & -0.00 & 0.91 \\
					\hline
				\multicolumn{7}{c}{Mediators $X$ (first part): study program, professional plans, intended industry and occupation}\\
				\hline
				program: business admin & 0.71 & 0.72 & 0.02 & 0.58 & -0.01 & 0.70 \\
				program: economics & 0.03 & 0.03 & 0.00 & 0.96 & 0.00 & 0.97 \\
				program: communication & 0.13 & 0.03 & -0.11 & 0.00 & 0.02 & 0.40 \\
				program: business IT & 0.07 & 0.19 & 0.13 & 0.00 & -0.00 & 0.93 \\
				future plans: work full time & 0.64 & 0.61 & -0.04 & 0.25 & 0.05 & 0.09 \\
				future plans: education & 0.44 & 0.44 & -0.00 & 0.96 & -0.03 & 0.39 \\
				industry: construction & 0.01 & 0.03 & 0.01 & 0.10 & -0.03 & 0.09 \\
				industry: trade and sales & 0.40 & 0.50 & 0.11 & 0.40 & 0.07 & 0.22 \\
				industry: transport and warehousing & 0.02 & 0.03 & 0.01 & 0.23 & -0.01 & 0.66 \\
				industry: hospitality and restaurants & 0.05 & 0.01 & -0.04 & 0.00 & -0.01 & 0.31 \\
				industry: information and communication & 0.38 & 0.31 & -0.08 & 0.02 & 0.03 & 0.36 \\
				industry: finance and insurance & 0.28 & 0.44 & 0.16 & 0.00 & 0.03 & 0.35 \\
				industry: consulting & 0.12 & 0.16 & 0.03 & 0.14 & -0.02 & 0.44 \\
				industry: education and science & 0.12 & 0.08 & -0.04 & 0.05 & -0.01 & 0.62 \\
				industry:health and social care & 0.06 & 0.04 & -0.02 & 0.17 & -0.01 & 0.71 \\
				occupation: general/strategic management & 0.25 & 0.37 & 0.12 & 0.00 & -0.01 & 0.71 \\
				occupation: marketing & 0.35 & 0.27 & -0.08 & 0.01 & -0.04 & 0.23 \\
				occupation: controlling & 0.10 & 0.15 & 0.05 & 0.03 & 0.00 & 0.89 \\
				occupation: finance & 0.20 & 0.29 & 0.10 & 0.00 & 0.01 & 0.59 \\
				occupation: sales & 0.07 & 0.11 & 0.04 & 0.03 & -0.02 & 0.31 \\
				occupation: technical/engineering & 0.05 & 0.11 & 0.06 & 0.00 & -0.04 & 0.14 \\
				occupation: human resources & 0.22 & 0.08 & -0.14 & 0.00 & 0.00 & 1.00 \\
				position: manager & 0.29 & 0.45 & 0.17 & 0.00 & 0.02 & 0.50 \\
					\hline
				\multicolumn{7}{c}{Mediators $X$ (second part): job expectations/views, preferences and plans concerning work and family life}\\
				\hline
				expect: well paid & 3.91 & 3.90 & -0.01 & 0.84 & -0.04 & 0.43 \\
				expect: invest in employees & 4.37 & 4.16 & -0.20 & 0.00 & -0.01 & 0.89 \\
				expect: good relations with boss & 4.54 & 4.25 & -0.29 & 0.00 & 0.10 & 0.08 \\
				expect: job security& 4.05 & 3.73 & -0.32 & 0.00 & 0.04 & 0.51 \\
				expect: family friendly & 3.60 & 2.81 & -0.78 & 0.00 & 0.03 & 0.69 \\
				expect: interesting tasks  & 3.98 & 3.80 & -0.18 & 0.01 & 0.04 & 0.50 \\
				expect: identification with work & 3.78 & 3.54 & -0.24 & 0.00 & 0.06 & 0.32 \\
				expect: priorities are flexible & 3.62 & 3.50 & -0.12 & 0.06 & -0.05 & 0.43 \\
			    views: fast decision making  & 2.94 & 3.04 & 0.10 & 0.08 & 0.01 & 0.80 \\
				views: competitive atmosphere & 2.87 & 3.20 & 0.33 & 0.00 & -0.06 & 0.31 \\
				views: self responsibility & 3.56 & 3.77 & 0.21 & 0.01 & 0.02 & 0.81 \\
				views:hierarchical structure & 2.70 & 2.94 & 0.24 & 0.00 & -0.06 & 0.43 \\
				stable partnership in 5-10 years & 0.79 & 0.71 & -0.08 & 0.01 & 0.02 & 0.58 \\
				preference for family & 0.22 & 0.32 & 0.10 & 0.00 & 0.01 & 0.64 \\
				preference for career & 0.02 & 0.06 & 0.03 & 0.01 & 0.00 & 0.92 \\
				wants children (0=no, 1=maybe, 2=yes)& 1.67 & 1.70 & 0.03 & 0.49 & -0.04 & 0.47 \\
				\hline
				\multicolumn{7}{c}{Outcomes $Y$}\\
				\hline
				gross wage category after studying & 6.01 & 7.16 & 1.15 & 0.00 &  &  \\
				gross wage category 3 yrs after studying & 8.57 & 10.04 & 1.48 & 0.00 &  &  \\
				\hline								
			\end{tabular}
		\end{center}
		\par
		Note: Trimming is 0.02 for the balancing tests after re-weighting, such that observations with propensity scores smaller than 0.02 or larger than 0.98 are dropped.}
\end{table}

The last 2 columns of Table \ref{descript1} provide mean differences across gender and p-values after reweighting treated observations by the inverse of the probit estimate of the propensity score $\Pr(G=1|X,W)$ and non-treated observations by the inverse of the estimate of $1-\Pr(G=1|X,W)$.\footnote{65 are dropped due to missing values in $W$, $X$, or $G$.} Such reweighting allows assessing whether the propensity score utilized in our IPW procedure (see Table \ref{resultstrim2} below) successfully balances differences in $W$ and  $X$ across gender as required for evaluating the explained and unexplained components. This indeed appears to be the case  when dropping observations with extreme propensity scores below 0.02 or above 0.98 (such that trimming is equal to 0.02 as in Table \ref{resultstrim2}), as most p-values are beyond statistical levels of significance. Only 3 differences in elements of $W$ or $X$ are statistically significant at the 10\% level, while no difference is significant at the 5\% level. Figure \ref{commonsupport} in the appendix displays the propensity score distributions separately for females and males and demonstrates that they decently overlap.

Table \ref{resultsob} provides the results of the Oaxaca-Blinder decomposition when considering two sets of mediators. In the first approach (`subset of mediators'), we only include those $X$ variables that are related to  characteristics typically observed and considered in decompositions, namely the study program,  job or educational plans after finishing the BA studies, as well as the intended industry, occupation, and job position. In our second approach, we in addition include variables that are typically not observed in data sets used for decompositions, but available in our questionnaire (`all mediators'): job expectations and views, as well as preferences and plans concerning work and family life. We report the indirect and direct effects (or explained and unexplained components) when considering either females  (`indir.f', `dir.f'), or - as it is common in wage decompositions - males (`indir.m', `dir.m') as reference group ($G=1$). In either case, the direct and indirect effects sum up to the total gap in wage expectations, defined as the mean difference between males and females (`total m-f'). The results are presented for two outcomes, namely the expected gross wage category after finishing the studies and 3 years later. Besides the point estimates (`est'), standard errors (`se') and p-values (`pval') based on 499 bootstrap replications are reported.

\begin{table}[!htbp]
	\caption{\label{resultsob}  Oaxaca Blinder decomposition} {\footnotesize
		\begin{center}
			\begin{tabular}{r|ccccc|ccccc}
				\hline\hline
				& \multicolumn{5}{c|}{subset of mediators} & \multicolumn{5}{c}{all mediators} \\
				& total m-f & indir.f & dir.f & indir.m & dir.m & total m-f & indir.f & dir.f & indir.m & dir.m \\
				\hline
				&  \multicolumn{10}{c}{outcome: gross wage category after studying}\\
				\hline
				est & 1.12 & 0.43 & 0.69 & 0.58 & 0.55 & 1.07 & 0.49 & 0.58 & 0.69 & 0.38 \\
				se & 0.16 & 0.15 & 0.22 & 0.16 & 0.22 & 0.16 & 0.19 & 0.23 & 0.19 & 0.23 \\
				pval & 0.00 & 0.00 & 0.00 & 0.00 & 0.01 & 0.00 & 0.01 & 0.01 & 0.00 & 0.10 \\
				 missings  &  & 45 &  &  &  &  & 72 &  &  &  \\
				\hline
				&  \multicolumn{10}{c}{outcome: gross wage category 3 yrs after studying}\\
				\hline
				est & 1.44 & 0.55 & 0.89 & 0.68 & 0.76 & 1.41 & 0.50 & 0.91 & 0.78 & 0.63 \\
				se & 0.18 & 0.16 & 0.23 & 0.16 & 0.22 & 0.19 & 0.22 & 0.27 & 0.21 & 0.26 \\
				pval & 0.00 & 0.00 & 0.00 & 0.00 & 0.00 & 0.00 & 0.02 & 0.00 & 0.00 & 0.02 \\
				missings   &  & 45 &  &  &  &  & 72 &  &  &  \\
				\hline
			\end{tabular}
		\end{center}
		\par
		Note: `total m-f'provides the total expectational wage gap between males and females. , `indir.f' and `dir.f' give the indirect (or explained) and direct (or unexplained) components when females are the reference group.  `indir.m' and `dir.m' give the respective components when males are the reference group.  Point estimates (`est') as well as standard errors (`se') and p-values (`pval') based on 499 bootstrap replications are reported.  `missings' provides the numbers of dropped observations due to missingness in variables.}
\end{table}

The expectational wage gap between males and females amounts to more than one category (=500 CHF) for wages after studying and to more than  1.4 categories 3 years later. The wage gap is driven by both observed characteristics $X$ and unexplained factors. Furthermore, when the set of variables in $X$ is extended from the subset to all mediators, the magnitude of the indirect effect (or explained component) increases and that of the direct effect (unexplained component) decreases. However, for either outcome, reference group, and definition of mediators, any direct and indirect effects are highly statistically significant, suggesting that even our atypically rich set of mediators cannot fully explain the expectational gender wage gap.

\begin{table}[!htbp]
	\caption{\label{resultstrim2}  Decomposition with trimming equal to 0.02} {\footnotesize
		\begin{center}
			\begin{tabular}{r|ccccc|ccccc}
				\hline\hline
				& \multicolumn{5}{c|}{ M1} & \multicolumn{5}{c}{M1+M2} \\
 & total  m-f & indir.f & dir.f & indir.m & dir.m & total m-f & indir.f & dir.f & indir.m & dir.m \\
     \hline
&  \multicolumn{10}{c}{outcome: gross wage category after studying}\\
\hline
est & 1.06 & 0.37 & 0.69 & 0.56 & 0.50 & 0.94 & 0.45 & 0.49 & 0.65 & 0.29 \\
se & 0.17 & 0.18 & 0.24 & 0.17 & 0.22 & 0.18 & 0.24 & 0.27 & 0.30 & 0.34 \\
pval & 0.00 & 0.04 & 0.01 & 0.00 & 0.02 & 0.00 & 0.06 & 0.07 & 0.03 & 0.39 \\
  missings / trimmed &  & 45 & / & 3 &  &  & 72 & / & 25 &  \\
    \hline
   &  \multicolumn{10}{c}{outcome: gross wage category 3 yrs after studying}\\
   \hline
   est & 1.37 & 0.50 & 0.87 & 0.67 & 0.71 & 1.29 & 0.48 & 0.81 & 0.81 & 0.48 \\
   se & 0.17 & 0.19 & 0.25 & 0.16 & 0.21 & 0.19 & 0.31 & 0.34 & 0.25 & 0.28 \\
   pval & 0.00 & 0.01 & 0.00 & 0.00 & 0.00 & 0.00 & 0.12 & 0.02 & 0.00 & 0.09 \\
   missings / trimmed &  & 45 & / & 4 &  &  & 72 & / & 20 &  \\
   \hline
\end{tabular}
\end{center}
\par
Note: `total m-f'provides the total expectational wage gap between males and females. , `indir.f' and `dir.f' give the indirect (or explained) and direct (or unexplained) components when females are the reference group.  `indir.m' and `dir.m' give the respective components when males are the reference group.  Point estimates (`est') as well as standard errors (`se') and p-values (`pval') based on 499 bootstrap replications are reported.  `missings' and `trimmed' provide the numbers of dropped observations due to missingness in variables and extreme propensity scores, respectively.}
\end{table}

Table \ref{resultstrim2} reports the results for IPW, which allows controlling for potential confounders $W$ and relaxing linearity assumptions. We to this end use the `medweight' command of the `causalweight' package by \citeasnoun{BodoryHuber2018} for the statistical software `R', with trimming set to 0.02 and 499 bootstrap replications for the estimation of the standard errors. Even though the precision of estimation is somewhat lower than before, the results are qualitatively similar to the Oaxaca-Blinder decomposition.
By and large, both the direct and indirect effects remain important for explaining the expectational wage gap. Using a larger trimming threshold of 0.04 results in quite stable results (see Table  \ref{resultstrim4} in the appendix).

\newpage
\section{Conclusion}\label{conclusion}

Using novel survey data from students from the Business School of the Bern University of Applied Science (BUAS) and the Faculty of Economic and Social Sciences of the University of Fribourg, this paper quantifies the role of career and family preferences on gender differences in wage expectations in Switzerland. After having documented the presence of gender differences in wage expectations, this paper provides three main results:
First, gender differences in wage expectations account for more than 1 salary class (CHF500) right after studying, and for more than 1.4 salary classes 3 years later. Both males and females overestimate actual wages from comparable graduates. Second, results from an information intervention (about median wages earned in Switzerland) suggest that only males (incorrectly) revise their expected wages three years after studying upward, by about 0.6 of a salary class (CHF300). This results in a larger expectational gender wage gap, by potentially exacerbating over-confidence. Third, the expectational wage gap is explained by both observed and unobserved factors. Using mediation analysis (which permits explicating endogeneity issues), we find that the expectational gender wage gap cannot be fully explained even after considering a very rich set of mediators, inclusive of career choices, family preferences and family plans. Results are stable under different specifications, including trimming thresholds and the randomization of the survey questionnaires.


%
%

\newpage
{\footnotesize
	
	\bibliographystyle{econometrica}
	\bibliography{expectations}}

\pagebreak

{\large \renewcommand{\theequation}{A-\arabic{equation}}
	\setcounter{equation}{0} \appendix }
\appendix \numberwithin{equation}{section}

\section{Appendix}

\subsection{Tables and figures}

\begin{center}
\begin{figure}[!htbp]
\caption{Common support for estimated $\Pr(G=1|W,X)$} \label{commonsupport}
\centering
\includegraphics[scale=0.7]{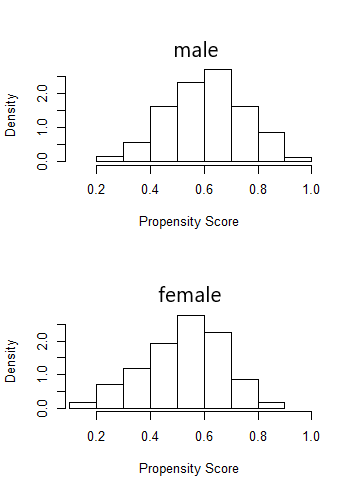}
\end{figure}
\end{center}

\begin{table}[!htbp]
	\caption{\label{resultstrim4}  Decomposition with trimming equal to 0.04} {\footnotesize
		\begin{center}
			\begin{tabular}{r|ccccc|ccccc}
				\hline\hline
				& \multicolumn{5}{c|}{M1} & \multicolumn{5}{c}{M1+M2} \\
				& total m-f & indir.f & dir.f & indir.m & dir.m & total m-f & indir.f & dir.f & indir.m & dir.m \\
				\hline
				&  \multicolumn{10}{c}{outcome: gross wage category after studying}\\
				\hline
				est & 1.05 & 0.37 & 0.68 & 0.55 & 0.50 & 0.90 & 0.35 & 0.55 & 0.64 & 0.26 \\
				se & 0.17 & 0.17 & 0.24 & 0.16 & 0.22 & 0.19 & 0.23 & 0.29 & 0.25 & 0.29 \\
				p-value & 0.00 & 0.03 & 0.00 & 0.00 & 0.03 & 0.00 & 0.12 & 0.06 & 0.01 & 0.36 \\
				missings / trimmed &  & 45 & / & 10 &  &  & 72 & / & 61 &  \\
				\hline
				&  \multicolumn{10}{c}{outcome: gross wage category 3 yrs after studying}\\
				\hline
				est & 1.36 & 0.49 & 0.86 & 0.66 & 0.70 & 1.18 & 0.48 & 0.71 & 0.72 & 0.46 \\
				se & 0.21 & 0.18 & 0.27 & 0.17 & 0.25 & 0.22 & 0.25 & 0.33 & 0.24 & 0.28 \\
				p-value & 0.00 & 0.00 & 0.00 & 0.00 & 0.00 & 0.00 & 0.06 & 0.03 & 0.00 & 0.10 \\
				missings / trimmed &  & 45 & / & 8 &  &  & 72 & / & 64 &  \\
				\hline
			\end{tabular}
		\end{center}
		\par
		Note: `total m-f'provides the total expectational wage gap between males and females. , `indir.f' and `dir.f' give the indirect (or explained) and direct (or unexplained) components when females are the reference group.  `indir.m' and `dir.m' give the respective components when males are the reference group.  Point estimates (`est') as well as standard errors (`se') and p-values (`pval') based on 499 bootstrap replications are reported. `missings' and `trimmed' provide the numbers of dropped observations due to missingness in variables and extreme propensity scores, respectively.}
\end{table}

\subsection{Questionnaire}



\section*{Supplementary material}

## A.2 Questionnaire

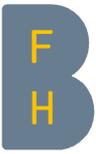

Dear students

The answers to the following questions are to be used for research. There is no identifier that could trace the questions to any particular individual, so your answers will remain strictly confidential.

The survey is fully voluntary. If you choose to answer it, know that I do appreciate your taking the time.

Thank you and best regards,
Ana Fernandes

v_p

| | |
|---|---|
| **A1** | **General questions about yourself** |
| A1.1 | How old are you? _____ |
| A1.2 | Your gender is: □ Female □ Male |
| A1.3 | What is your nationality (indicate the country you feel closer to, if multiple are possible)? _____ |
| A1.4 | Which degree programme are you taking at the Bern University of Applied Sciences?<br><br>□ <u>Bachelor of</u> _____ I □ <u>Master of</u> _____ I □ _____ |
| A1.5 | Which study model have you chosen?<br><br>□ <u>Full-time degree programme</u> I □ <u>Part-time degree programme</u> I □ _____ |
| **A2** | **Questions about your professional path** |
| | 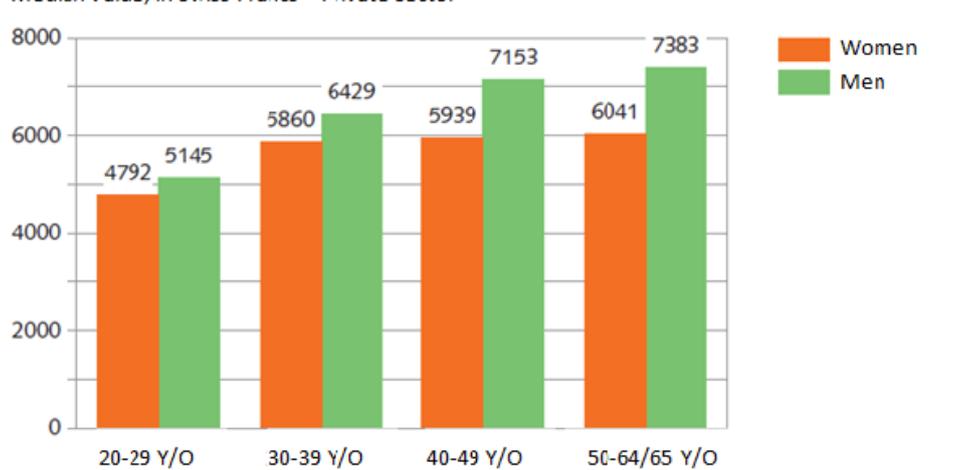 |
| A2.1 | Which statement best describes you at the moment?<br><br>□ You already have work experience (full-time or part-time)<br><br>□ You have held odd jobs to supplement income or not worked at all<br><br>□ Other _____ |
| A2.2 | What are your plans regarding work or study once you complete the degree you are currently enrolled in? **Multiple answers possible**:<br><br>□ Work full-time □ Work part-time □ Further education (MA, or another advanced degree, or...)<br><br>□ Do not intend to seek employment □ Other _____ |

| | |
|---|---|
| A2.3 | Consider the following company and job attributes. Please assign a grade from 1 to 5 to each one, with 5 being highly valued and 1 being not important.<br><br>__ Company pays well<br><br>__ Company invests in people and in their careers<br><br>__ Flexible working conditions with respect to schedule and location<br><br>__ Pleasant relations with boss and colleagues<br><br>__ Job security<br><br>__ Supportive policies to accommodate maternity/paternity<br><br>__ Daily tasks are intellectually stimulating<br><br>__ Your work in the company contributes to a cause that is important to you<br><br>__ Flexibility in setting priorities and tasks<br><br>__ The commuting time between your home and the office is short |
| A2.4 | Consider the following company and job attributes. Please assign a grade from 1 to 5 to each one, with 5 being very good and 1 being very bad.<br><br>__ Long hours<br><br>__ Repetitive<br><br>__ Need to systematically make decisions in a very short time<br><br>__ Mistakes would impact negatively the well-being of many people<br><br>__ The environment is very competitive<br><br>__ Personal accountability for successes and mistakes<br><br>__ Hierarchical firm structure |
| A2.5 | What are your expected monthly gross earnings (no taxes deducted) when you graduate from the current studies (in Swiss Francs)?<br><br>□ Below 3'500   □ 3'500-4'000   □ 4'000-4'500   □ 4'500-5'000   □ 5'000-5'500   □ 5'500-6'000<br><br>□ 6'000-6'500   □ 6'500-7'000   □ 7'000-7'500   □ 7'500-8'000   □ 8'000-8'500   □ 8'500-9'000<br><br>□ 9'000- 9'500   □ 9'500-10'000   □ 10'000-10'500   □ 10'500-11'000   □ More than 11'000 |
| A2.6 | What are your expected monthly gross earnings (no taxes deducted) 3 years after graduation (in Swiss Francs)?<br><br>□ Below 3'500   □ 3'500-4'000   □ 4'000-4'500   □ 4'500-5'000   □ 5'000-5'500   □ 5'500-6'000<br><br>□ 6'000-6'500   □ 6'500-7'000   □ 7'000-7'500   □ 7'500-8'000   □ 8'000-8'500   □ 8'500-9'000<br><br>□ 9'000- 9'500   □ 9'500-10'000   □ 10'000-10'500   □ 10'500-11'000   □ More than 11'000 |

| | | |
|---|---|---|
| A2.7 | Set your salary at graduation at 100. How much do you think will be earned by a student with the same characteristics as yours but of the other gender?<br><br>☐ 50  ☐ 60  ☐ 70  ☐ 80  ☐ 90  ☐ Salary at 100  ☐ 110  ☐ 120  ☐ 130  ☐ 140  ☐ 150<br>Graduation<br><br>If possible, please say why_____ | |
| A2.8 | Set your salary *10 years after* graduation at 100. How much do you think will be earned by a student with the same characteristics as yours but of the other gender?<br><br>☐ 50  ☐ 60  ☐ 70  ☐ 80  ☐ 90  ☐ Salary at 100  ☐ 110  ☐ 120  ☐ 130  ☐ 140  ☐ 150<br>Graduation<br><br>If possible, please say why _____ | |
| A2.9 | In which industry do you expect to be working upon completion of your degree?<br><br>☐ Agriculture, forestry, fishing, mining  ☐ Manufacturing  ☐ Construction  ☐ Trade and Sales<br><br>☐ Transportation and storage  ☐ Accommodation and food service  ☐ Information and communication<br><br>☐ Finance and insurance  ☐ Real estate  ☐ Consulting  ☐ Public Administration<br><br>☐ Education and Science  ☐ Health care and social care  ☐ Arts, entertainment and recreation<br><br>☐ Other _____  ☐ Don't know | |
| A2.10 | In which occupation do you expect to be working upon completion of your degree?<br><br>☐ General and strategic management  ☐ Marketing  ☐ Controlling  ☐ Finance<br><br>☐ Research and Development  ☐ Sales  ☐ Logistics  ☐ Production<br><br>☐ Technical support and engineering  ☐ HR  ☐ Unskilled occupation<br><br>☐ Other _____  ☐ Don't know | |
| A2.11 | In your future professional career, which role would you prefer to have:<br><br>☐ A management/leadership position in a company (for example CEO)?<br><br>☐ A specialist/supporting position (for example the head of a cabinet providing specialized support to the CEO)?<br><br>☐ Not sure yet<br><br>☐ If you chose one of the first two options above, please explain why you prefer one over the other<br>_____ | |
| A3 | **Questions about family and other personal dimensions** | |
| A3.1 | In the next 5 to 10 years, do you see yourself in a stable partner relationship?<br><br>☐ Yes  ☐ No  ☐ Not sure | |

| | |
|---|---|
| A3.2 | Would you like to have children eventually? If your answer is "yes," please indicate how many.<br><br>□ No    □ Yes    □ Not sure    (If yes) How many? _____ |
| A3.3 | If you were in a stable partner relationship with children in the next 5 to 10 years, which employment type would you prefer:<br><br>□ Work full-time    □ Work part-time    □ Not work    □ Other _____ |
| A3.4 | If you were in a stable partner relationship with children in the next 5 to 10 years, which employment type would you prefer your partner to have:<br><br>□ Work full-time    □ Work part-time    □ Not work    □ The employment status preferred by the partner<br><br>□ Other_____ |
| A3.5 | When you were a young child of pre-Kindergarten age (roughly ages 0-4), what was the employment status of your parents in general?<br><br>□ My mother did not work           □ My father did not work<br><br>□ My mother worked part-time      □ My father worked part-time<br><br>□ My mother worked full-time      □ My father worked full-time<br><br>□ Other: _____            □ Other: _____ |
| A3.6 | When you were a young child of Kindergarten age (roughly ages from 4-6), what was the employment status of your parents in general?<br><br>□ My mother did not work           □ My father did not work<br><br>□ My mother worked part-time      □ My father worked part-time<br><br>□ My mother worked full-time      □ My father worked full-time<br><br>□ Other: _____            □ Other: _____ |
| A3.7 | During your first school years (1st to 6th grade), what was the employment status of your parents in general?<br><br>□ My mother did not work           □ My father did not work<br><br>□ My mother worked part-time      □ My father worked part-time<br><br>□ My mother worked full-time      □ My father worked full-time<br><br>□ Other: _____            □ Other: _____ |
| A3.8 | Which sentence describes you best?<br><br>□  Family life is important to me<br><br>□  Family life is important to me but so is my professional career<br><br>□  I value my professional career more than I value my family life |

| A3.9 | Please tell us about your family members by indicating the number of relatives that you have of each type below.<br><br>Siblings ____     Living parents or step parents ____     Living grandparents ____<br><br>Partner/girlfriend/boyfriend ____     Own children ____ |
|---|---|
| A3.10 | Please answer this question if you have siblings. How many of your siblings are older than you?<br><br>I have ___ sisters who are older than me.     I have ___ brothers who are older than me. |
| A3.11 | How many people do you share your place of residence with? _____<br><br>If you do not live alone, please specify who you are living with (parents, siblings, ...)<br><br>_____ |
| A3.12 | What type of residence do you live in?<br><br>□ Rented flat     □ Rented house     □ Own flat     □ Own house     □ Other _____ |
| A3.13 | What is the highest educational attainment of your parents:<br><br>|  | Mother | Father |<br>|---|---|---|<br>| Compulsory schooling (up to lower secondary schooling) or less | □ | □ |<br>| Apprenticeship or vocational degree | □ | □ |<br>| Upper secondary schooling (including high school) | □ | □ |<br>| University (Bachelor degree) | □ | □ |<br>| University (Masters degree) | □ | □ |<br>| University (PhD/doctorate) | □ | □ |<br>| Another (please specify) | □ | □ |<br>|  | _____ | _____ | |
| A3.14 | How would you evaluate your material well-being compared to the average person in Switzerland? I am doing:<br><br>| Much worse than average person | Worse than average person | Roughly the same as average person | Better than average person | Much better than average person |<br>|---|---|---|---|---|<br>| □ | □ | □ | □ | □ | |

6



\end{document}